\documentclass{pramana}

\usepackage{graphicx,amsmath,bm}

\begin{document}

\title{High-sensitivity measurement of Rydberg population via two-photon excitation in atomic vapour using optical heterodyne detection technique}


\author{Arup Bhowmick\textsuperscript{1,*}, Dushmanta Kara\textsuperscript{1}\and Ashok K. Mohapatra\textsuperscript{1}}
\affilOne{\textsuperscript{1} School of Physical Sciences, National Institute of Science Education and Research Bhubaneswar, HBNI, PO-Jatni, 752050, India}


\twocolumn[{

\maketitle

\corres{arup.b@niser.ac.in}

\msinfo{1 January 2015}{1 January 2015}{1 January 2015}

\begin{abstract}
We demonstrate a technique based on optical heterodyne detection to measure Rydberg population in thermal atomic vapour. The technique used a probe beam far off resonant to the D2 line of rubidium along with a reference beam with frequency offset by 800 MHz in the presence of a coupling laser that couples to Rydberg state via two-photon resonance. The polarizations of the probe, reference and coupling beams are suitably chosen such that only the probe beam goes through a non-linear phase shift due to two-photon process which is measured relative to the phase shift of the reference beam using optical heterodyne detection technique. We show that the technique has a sensitivity to measure the minimum phase shift of the order of few $\mu$rad. We have used a suitable model of two-photon excitation of a 3-level atom to show that the minimum phase shift measured in our experiment corresponds to Rydberg population of the order of $10^{-5}$. The corresponding probe absorption for the given laser parameters is of the order of $10^{-7}$. We demonstrate that this technique is insensitive to polarization impurity or fluctuations in the beams. The technique is particularly useful in measuring Rydberg population via two-photon excitation in thermal vapour where multi channel plates (MCP) could be relatively difficult to impliment. It can also be used in ultra-cold atomic sample with suitable laser parameters.
\end{abstract}



}]


\doinum{12.3456/s78910-011-012-3}
\artcitid{\#\#\#\#}
\volnum{123}
\year{2016}
\pgrange{23--25}
\setcounter{page}{23}
\lp{25}

\section{Introduction}
Rydberg atoms are enriched with enhanced many body interactions. When atoms in a dense frozen ensemble are excited to the Rydberg state using narrow band lasers, strong Rydberg-Rydberg interactions lead to excitation blockade. This blockade interaction generates a highly entangled many-body quantum state in an ensemble of atoms which has become the basis for fundamental quantum gates using atoms~\cite{jaks00,luki01,isen10,wilk10} or photons~\cite{frie05} and for realization of single photon source~\cite{saff02,dudi12}. Rydberg blockade in ultra-cold atoms and Bose-Einstein condensate (BEC) has been proposed~\cite{weim08,pohl09,pupi10,henk10} and observed~\cite{ange16} in magneto optical trap~\cite{tong04,sing04,cube05,vogt06,heid07,rait08,urba09,gaet09,dudin12,scha12,prit10,peyr12,firs13,pari12,webe15,jau16}. It has significant application to simulate many-body phenomenon~\cite{webe15} and the generation of entangled state~\cite{cano14} which are essential in the field of quantum engineering~\cite{labu14}. It is important to develop a detection mechanism for Rydberg excitation. A direct absorption of the probe in 2-photon resonance technique has been used for ultra-cold atoms to measure Rydberg population and the precision of the measurement is found to be $10^{-2}$~\cite{karl15}. A state selective spectroscopic detection of highly excited Rydberg population has also been reported~\cite{sand85}. Rydberg tomography of ultra-cold has been established to image Rydberg blockade~\cite{vala13}. Study of Rydberg excitation in thermal atomic vapor has been reported in a number of literature~\cite{balu13,koll12,zhan15,moha07,moha08,carr13} in the context of development of quantum engineering using the system. In the experiments with cold atomic ensemble, MCPs are normally used to detect Rydberg atoms after ionisation using a dc electric field which could be an issue with thermal vapor experiments due to presence of charges sticking to the dielectric surface. Though, a recent experiment with electrically contacted thermal vapor cell was used~\cite{barr13}, all optical techniques are relatively easier to study Rydberg excitation in thermal vapor. Rydberg excitation using all optical techniques are already demonstrated in electromagnetically induced transparency (EIT) regime in thermal vapor cell~\cite{moha07,bhow16,akul04} and in micron sized vapor cell~\cite{kubl10}. Also, Rydberg excitation with two and three photon excitation have been studied in large probe absorption regime~\cite{firs16,carr13}. In this article, we present a method based on optical heterodyne detection technique to measure the dispersion of a probe beam due to two photon excitation to Rydberg state. Heterodyne detection technique has been used to study absorption and dispersion of coherent two-photon transition in an atomic ensemble~\cite{mull96}, Zeeman coherence induced anomalous dispersion~\cite{akul99}, and enhanced Kerr nonlinearity in two-level atoms~\cite{akul04}. The technique has also been used to measure the XPM of a probe and a control beam in an N system using cold atoms~\cite{kang03,lo10,han08} and also Rydberg EIT~\cite{bhow16}. We show with a suitable model that Rydberg population can be measured directly from the dispersion in strong probe regime. We also show that the technique is insensitive to the small imperfections and fluctuations in the polarizations and intensity of the coupling lasers. The technique is sensitive enough to measure probe absorption of the order of $10^{-7}$. 
\begin{figure}[t]
\includegraphics[angle=0,scale=0.3]{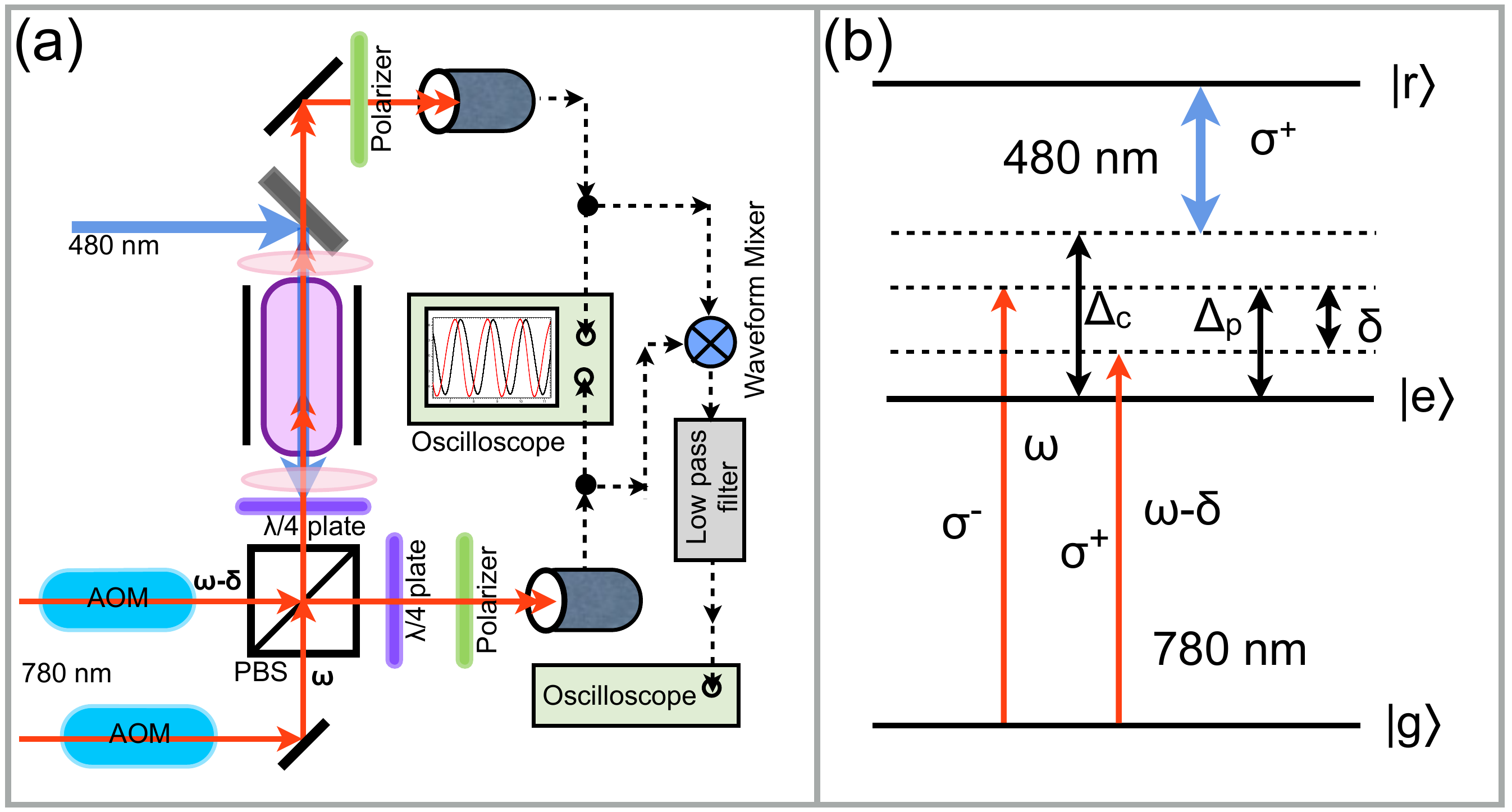}
\caption{(a) Schematic of the experimental set up. (b) Energy level diagram for 2-photon transition to the Rydberg state. Two probe beams with $\sigma^+$ and $\sigma^-$ polarizations couple the transition $5$S$_{1/2}$, F$=3$ $\left(\left|g\right\rangle\right)$ $\longrightarrow$ $5$P$_{3/2}$ $\left(\left|e\right\rangle\right)$ of $^{85}$Rb. The coupling laser with $\sigma^+$ polarization couples the transition $5$P$_{3/2}$ $\left(\left|e\right\rangle\right)$ $\longrightarrow$ $n$S$_{1/2}$ $\left(\left|r\right\rangle\right)$. The probe (coupling) detuning is $\Delta_{p}$ ($\Delta_{c}$) and frequency offset between the probe beams is $\delta$.}
\label{fig1}
\end{figure}

\section{Experimental methodology}
The schematic of the experimental set up is shown in figure 1(a). A technique based on optical heterodyne detection is used to measure the dispersion of a probe beam due to two-photon excitation to Rydberg state. The experimental set up is similar to our earlier experiment to study the optical non-linearity of Rydberg-EIT in thermal vapor~\cite{bhow16}. A probe beam along with a reference beam were derived from an external cavity diode laser operating at $780$ nm. A frequency offset of $800$ MHz was introduced between the probe and the reference beams by passing them through two acousto-optic modulators. Both the beams were made to superpose using a polarizing cube beam splitter (PBS). The interference beat signals were detected using two fast photo-detectors by introducing polarizers at both the output ports of the PBS. A controlled relative phase can be introduced between both the beat signal by using an optical phase shifter as explained in reference~\cite{bhow16}. The probe beams coming out of one of the output ports of the PBS propagate through a magnetically shielded rubidium vapor cell. The temperature of the vapor cell can be varied from $30$ $^0$C to $130$ $^0$C using a heater which was controlled by a PID-controller. With this range of temperature, the density of $^{85}$Rb can be varied in the range of $1.7\times 10^{10}$ cm$^{-3}$ - $3.0\times 10^{13}$ cm$^{-3}$. The coupling beam was derived from a frequency doubled laser system operating at wavelength in the range of $478 - 482$ nm and it counter-propagates the probe beams through the cell. The coupling and the probe beams were focused inside the cell using suitable lenses. The waist and the Rayleigh range of the probe (coupling) beams are $35$ $\mu$m ($50$ $\mu$m) and $12$ mm ($10$ mm), respectively. The peak Rabi frequencies of the laser beams and their variations over the length of the vapor cell were calculated~\cite{rabi} using the same parameters of the beams and were included in the theoretical model to fit all the experimental data.

The probe and the reference beams propagating through the medium can undergo different phase shift by choosing suitable polarizations of the probe, reference and the coupling beams. The polarization of the coupling beam was chosen to be $\sigma^{+}$. The reference beam with $\sigma^{+}$ polarization can not couple to the transition, $5$S$_{1/2}$ $\rightarrow$ $n$S$_{1/2}$ and doesn't go through any phase shift due to two-photon process. However, the probe beam with $\sigma^{-}$ polarization can couple the same two-photon transition and hence, goes through a phase shift due to two-photon excitation to the Rydberg state. This additional phase shift of the probe beam appears as a phase shift in the respective beat signal and is measured by comparing to the phase of the reference beat signal detected at the other output port of the PBS. Since, both the beat signals are the output of the same interferometer, the noise due to vibration and acoustic disturbances are strongly suppressed. The signal-to-noise ratio was further improved by using a lock-in amplifier. A mechanical chopper was used to modulate the intensity of the coupling beam at a rate of $6$ kHz 
which is used as reference to the lock-in amplifier. 

The output signal observed in the experiment is $S_L \propto\Re\left(\chi_{3L}\right)$ where $\chi_{3L}$ is the susceptibility of the probe field due to the two-photon process~\cite{bhow16}. In the experiment, the probe beam was stabilized at $1.3$ GHz blue detuned to the D2 line of $^{85}$Rb. With this detuning, absorption of the probe beam is negligible due to interaction with the D2 line at $130^0$C. The coupling laser frequency was scanned to observe the dispersion of the signal probe beam by measuring its phase shift due to the two-photon excitations to the Rydberg state.  We have observed that the simple model of a three-level atom interacting with a single probe and coupling beams matches well with the measurement using this technique if the frequency offset is much larger than the reference Rabi frequency~\cite{bhow16}. Hence, a large frequency offset of $800$ MHz was used in the experiment.

\begin{figure}[t]
\includegraphics[angle=0,scale=0.24]{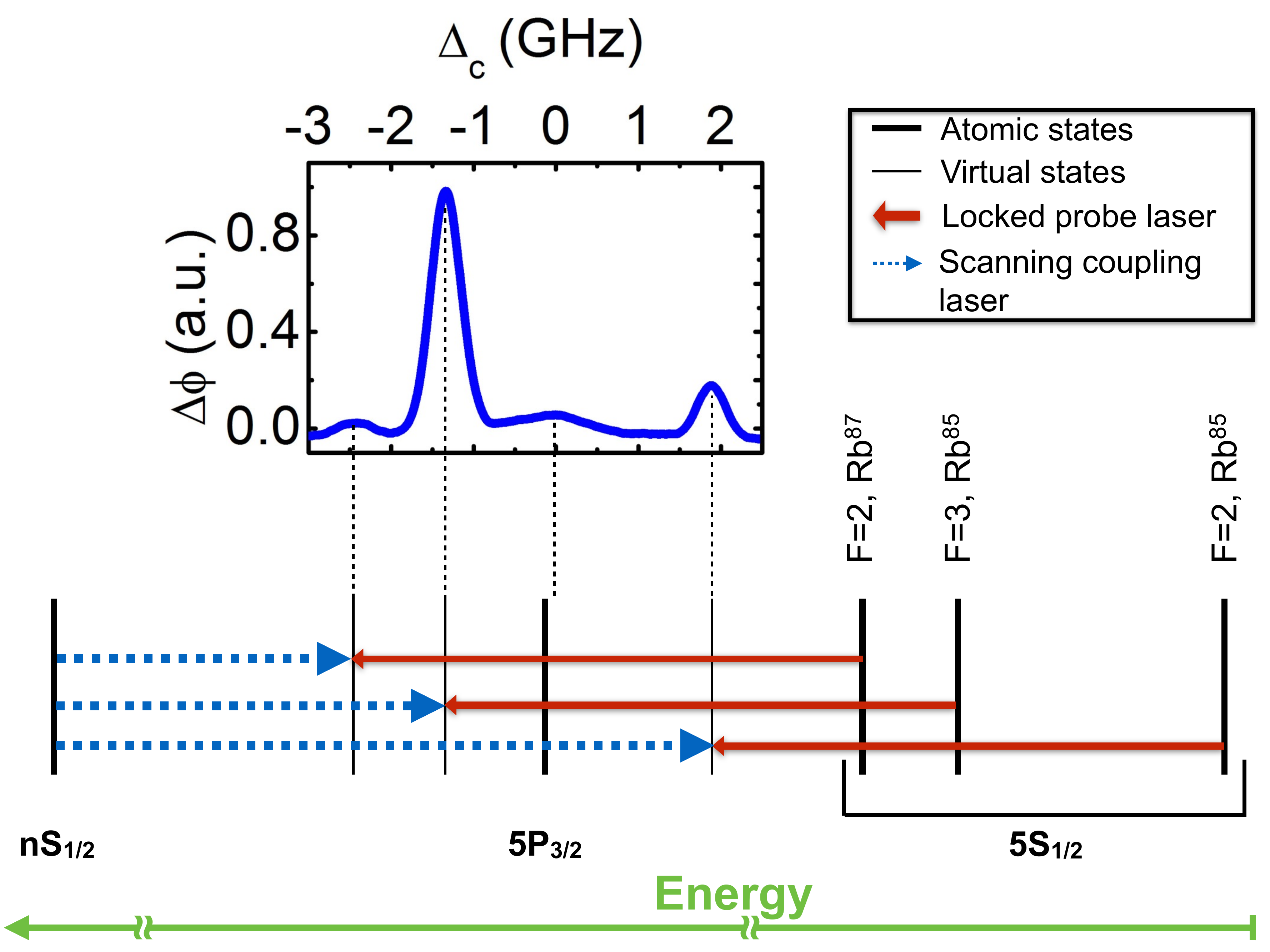}
\caption{Typical dispersion spectrum observed in the experiment by scanning coupling laser over $5$ GHz. The relevant energy levels of rubidium and respective dispersion peaks are depicted.}
\label{fig2}
\end{figure}

\begin{figure}[t]
\includegraphics[angle=0,scale=0.34]{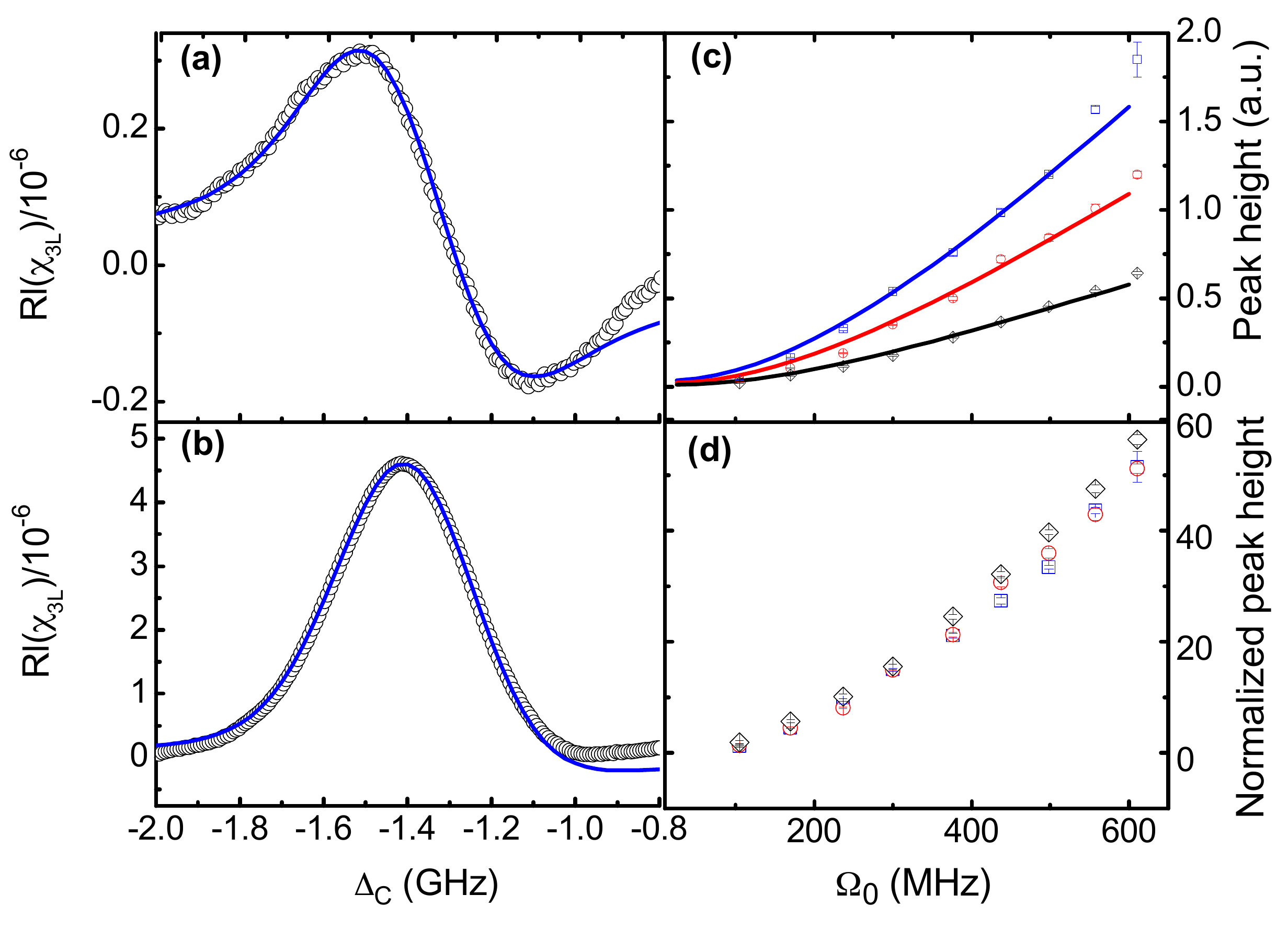}
\caption{Refractive index of the signal probe beam propagating through the medium with coupling scanning over the 2-photon resonance. The peak Rabi frequencies of the coupling beam was $24$ MHz and of the probe was (a) $60$ MHz and (b) $400$ MHz. (c) Measured dispersion peak height as a function of peak Rabi frequency of probe while coupling to the Rydberg state $n=33$ with atomic vapor densities $2.5\times 10^{12}$ cm$^{-3}$ $({\diamond})$, $1.25\times 10^{13}$ cm$^{-3}$ $({\circ})$ and $3.0\times 10^{13}/$ cm$^{-3}$ $({\square})$. The solid lines are the fitting using the theoretical model and a multiplication factor is used as the only fitting parameter which can be accounted for the overall gain in the experiment. (d) Dispersion peak height normalized to the peak height of a weak probe beam.}
\label{fig3}
\end{figure} 

\section{Theory}
The typical dispersion spectrum of the probe beam due to two-photon process is shown in figure \ref{fig2}. Two-photon resonance peaks corresponding to the transitions, $5$s$_{1/2}$F$=3\longrightarrow$ $n$s$_{1/2}$ and $5$s$_{1/2}$F$=2\longrightarrow$ $n$s$_{1/2}$ of $^{85}$Rb and $5$s$_{1/2}$F$=2\longrightarrow$ $n$s$_{1/2}$ of $^{87}$Rb are observed while the coupling beam is scanned over $5$ GHz. The dispersion peak corresponding to the $5$s$_{1/2}$F$=3\longrightarrow$ $n$s$_{1/2}$ transition of $^{85}$Rb was analyzed for the further study of Rydberg excitation. An usual dispersion profile of the two-photon resonance is observed for a weak probe beam as shown in  figure~\ref{fig3}(a). However, an absorptive like dispersion profile is observed for a stronger probe beam as shown in figure~\ref{fig3}(b). In order to model the observed dispersion spectrum in the experiment, we consider a Hamiltonian of a three-level atom interacting with a probe and a coupling beams. Using rotating wave approximation (RWA) in a suitable rotating frame, $H=-\frac{\hbar}{2}[(\Omega_{p} |e\rangle \langle g|+\Omega_{c} |r\rangle \langle e|)+h.c.]-\hbar [(\Delta_{p} - k_{p} v)|e\rangle \langle e| + (\Delta_{2} +\Delta k v)|r\rangle \langle r|]$. Where $\Omega_{p}$ ($\Omega_{c}$) is the probe (coupling) Rabi frequency and $\Delta_{p}$ ($\Delta_{c}$) is the probe(coupling) laser detuning with the two-photon detuning being $\Delta_{2} =\Delta_{p} +\Delta_{c}$. $k_{p}$ ($k_{c}$) is the wave vector of probe (coupling) laser, $\Delta k=k_{c} -k_{p}$ and  $v$ is the velocity of the atom. The density matrix equation of the system, $i\hbar\dot{\hat{\rho}}=[\hat{H},\hat{\rho}]+i\hbar\mathcal{L}_D\hat{\rho}$ are solved numerically in steady state. Here, $\mathcal{L}_D$ is the Lindblad operator which takes care of the decoherence in the system. In the numerical calculation, we have used the decay rates of the channels $\left|e\right\rangle\rightarrow\left|g\right\rangle$ and $\left|r\right\rangle\rightarrow\left|e\right\rangle$ as $2\pi\times 6$ MHz and $2\pi\times 0.01$ MHz, respectively. A population decay rate of the Rydberg state to the ground state denoted as $\Gamma_{rg}$ is used to account for the finite transit time of the thermal atoms in the laser beams. The dipole $\rho_{rg}$ dephase at a rate of $\frac{\Gamma_{rg}}{2}+\gamma_{rel}$, where $\gamma_{rel}$ accounts for the relative laser noise between the probe and the coupling beams. $\Gamma_{rg}$ and $\gamma_{rel}$ are of the order of $2\pi\times 0.5$ MHz. The susceptibility of the probe beam due to two-photon process averaged over the thermal motion of the atoms can be calculated as, 
\begin{equation}
\chi_{3L}=\frac{2N\left|\mu_{eg}\right|^2}{\epsilon_0\hbar\Omega_{p}}\frac{1}{\sqrt{2\pi}v_p}\int^{\infty}_{-\infty}\rho^{(3L)}_{eg}e^{-\frac{v^2}{2v^2_p}}dv
\end{equation}
 where $v_p$ is the most probable speed of the atoms, $N$ is the density and $\mu_{eg}$ is the dipole moment of the transition 
$\left|g\right\rangle\longrightarrow\left|e\right\rangle$. $\rho^{(3L)}_{eg}=\rho_{eg}-\rho^{(2L)}_{eg}$ where $\rho^{(2L)}_{eg}$ is due to the interaction of the probe beam with the 
$\left|g\right\rangle\longrightarrow\left|e\right\rangle$ transition in the absence of the coupling beam. The dispersion of the probe beam $\Re\left(\chi_{3L}\right)$ calculated using the 
above theoretical model can be compared with the experimental data.
In a further study, the variation of the dispersion peak height was measured as a function of probe Rabi frequency. The change in probe laser power 
resulted in a change in the beat signal amplitude. An RF attenuator at the output of the detectors was used to keep the amplitude of the beat signals constant irrespective of the probe laser power. 
The variation of the dispersion peak height as a function of probe Rabi frequency for different vapor densities is shown in figure~\ref{fig3}(c). The dispersion peak height calculated by the theoretical model 
for the same laser parameters and vapor density agrees well as shown in the figure~\ref{fig3}(c). When the peak height data is normalized with the dispersion peak of a weak probe beam, then all the data corresponding to different densities fall on the same line as shown in figure~\ref{fig3}(d). This observation suggest that the refractive index of the medium depends linearly on the vapor density and the Rydberg-Rydberg interaction has negligible effect.\\

\begin{figure}[t]
\includegraphics[angle=0,scale=0.35]{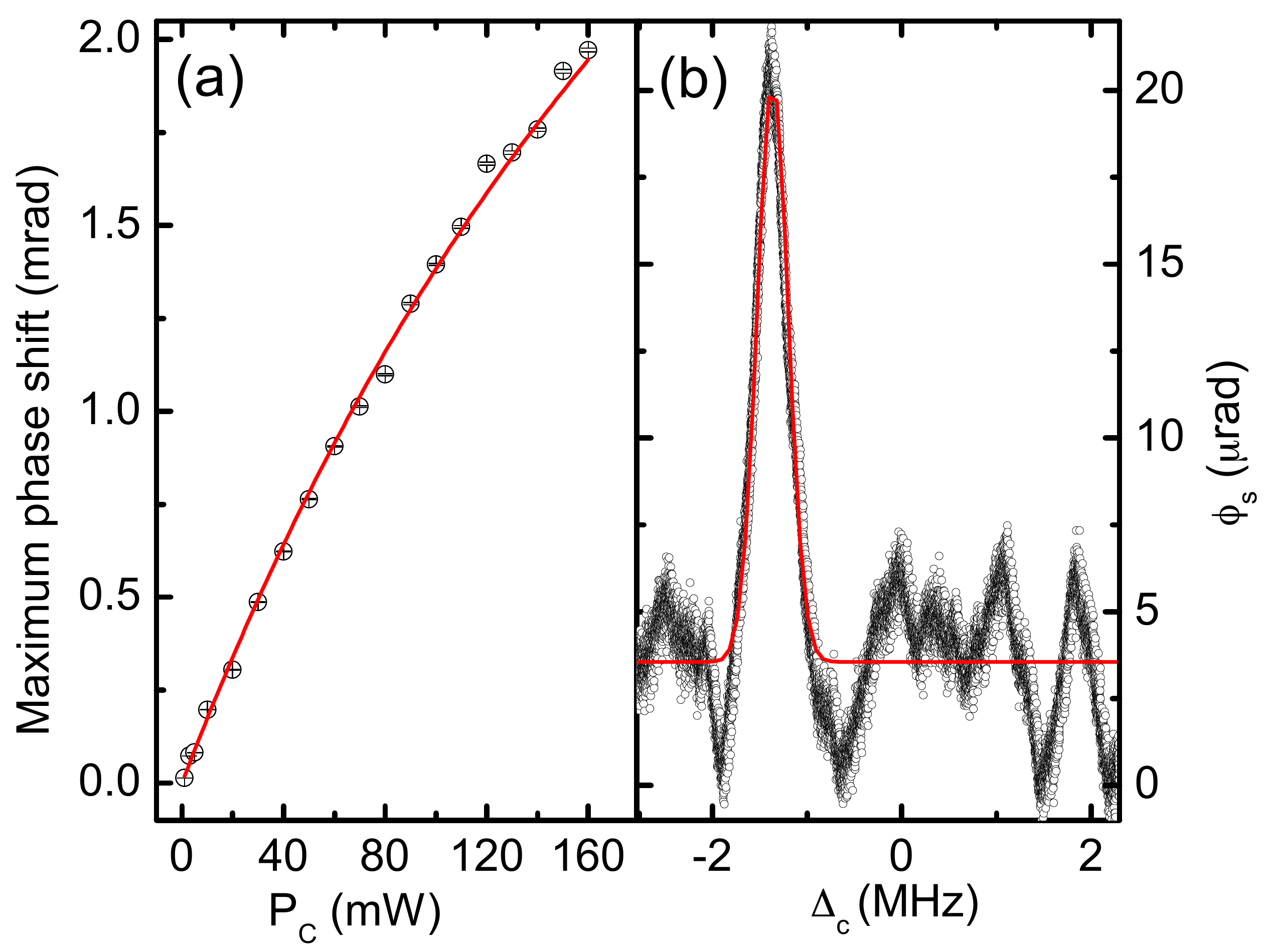}
\caption{(a) Phase shift of the probe at the peak of the dispersion spectrum as a function of the coupling power. Open circles correspond to the measured phase shift using heterodyne detection 
technique and the solid line is the fitting using the model discussed in the text.  (b) Observed dispersion spectrum by scanning the coupling laser over the $5S_{1/2} (F=3) \rightarrow nS_{1/2}$ 
transition of $^{85}$Rb with coupling laser power being $1$ mW. The signal is fitted with a gaussian function to determine the phase shift at the peak of the spectrum.}
\label{fig4}
\end{figure}

To get a further insight to the observed shape of the dispersion profile, we use the approximation $\Delta_p>>\Gamma_{eg},\Omega_p$ 
such that the intermediate state population is negligible. Then, in a regime $\Omega_p>>\Gamma_{eg}$, we find $\Re\left(\rho_{eg}\right)=
-\frac{\Omega_p}{2\left(\Delta_p-k_pv\right)} +\frac{1}{2\left(\Delta_p-k_pv\right)}\left[\Omega_p\rho_{rr}-\Omega_c\Re\left(\rho_{rg}\right)\right]$.
The first term in the above equation gives the susceptibility of probe interacting with $|g\rangle\longrightarrow|e\rangle$ transition in the absence of the coupling laser. Then the dispersion due 
to two-photon process is then
\begin{equation}
\Re\left(\rho^{(3L)}_{eg}\right)= \frac{1}{2\left(\Delta_p-k_pv\right)}\left[\Omega_p\rho_{rr}-\Omega_c\Re\left(\rho_{rg}\right)\right]
\end{equation}
$\rho_{rr}$ and $\rho_{rg}$ can be determined by approximating the 3-level atom as an effective 2-level atom by adiabatically eliminating the intermediate state~\cite{han13} for the large single photon detuning. In the regime,  $\Omega_p >> \Omega_c$, neglecting the second term in the above equation leads $\Re\left(\rho^{(3L)}_{eg}\right)\propto \rho_{rr}$ and hence the dispersion gives rise to an absorptive like profile. For small $\Omega_p$, contribution of the second term of the same equation gives a usual feature of dispersion profile as observed in the experiment. Since the dispersion is proportional to Rydberg population in a strong probe regime, this technique gives the direct information about the Rydberg excitation and hence becomes an all optical method to  detect the Rydberg atoms without the use of multi channel plate. To measure the sensitivity of the technique, we recorded the dispersion signal with different coupling Rabi frequency as shown in fig.(\ref{fig4}). Using the effective two-level atom, Rydberg population can be determined as  $\rho_{rr}=\frac{\Omega^{2}_{eff}}{\Delta^{2}_{eff} +2\Omega^{2}_{eff}+\Gamma^{2}_{rg}}$ where $\Omega_{eff}=\frac{\Omega_p\Omega_c}{2(\Delta_{p} -\Delta_{c})}$ and $\Delta_{eff}=2(\Delta_p +\Delta_c )+2\Delta kv +\frac{\Omega^{2}_{p}}{(\Delta_{p}- \Delta_{c} )}$ which are the effective Rabi frequency and detuning of the approximated two-level atom respectively. While the lasers are tuned within the Doppler width of the two-photon transition, the atoms with the velocity class resonant to the lasers contributes maximum to the Rydberg population. Using equation (1) and (2), the peak of the dispersion due to two-photon process can be approximated as $\Re\left({\chi_{3L}}\right)\approx\frac{\Omega_{eff}}{2\sqrt{\pi}\Delta kv_p}\frac{N\left|\mu_{eg}\right|^2}{\epsilon_0\hbar\Delta_p}\rho_{rr}$. Here, we have assumed that the peak of the dispersion corresponds to the atoms with zero velocity class within the effective Rabi frequency $\left(\Omega_{eff}/\Delta k\right)$ interacting resonantly with the lasers.

\section{Precision of the measurement}
The phase shift $(\phi_s)$ due to the two-photon process measured using the heterodyne detection technique is related to the susceptibility as $\Re\left(\chi_{3L}\right)=2\phi_s/k_pl$ where $l$ is the length of the vapor cell. $\phi_s$ can be determined from the experimental set up as $\phi_s=\frac{V_o}{G\eta_m}$ where $V_o$ and $G$ are the output and overall gain of the lock-in amplifier used in the experiment respectively. $\eta_m$ is the sensitivity of the RF mixer which was determined by using two RF signals with a known phase difference with same amplitudes as the beat signals of the heterodyne set up in the experiment. The phase shift at the peak of the dispersion was measured using the heterodyne detection technique by varying the coupling laser power is shown in figure~\ref{fig4}(a). The data were fitted with a simple formula which can be derived using above analysis. The dispersion spectrum with very weak coupling power is shown in figure~\ref{fig4}(b) and fitting with a Gaussian function gives the peak of the phase shift. Rydberg population can be determined from the measured phase shift as, $\rho_{rr}=\frac{2\epsilon_{0}\hbar \Delta_{p}}{N|\mu_{eg}|^{2}}\frac{\Delta k v_{p}}{\sqrt{\pi}\Omega_{eff}} \frac{\lambda_{p}}{l}\phi_{s}$. The Rydberg population at the peak of the spectrum shown in~\ref{fig4}(b) is found to be $\rho_{rr} \approx 10^{-5}$.  

\begin{figure}[t]
\includegraphics[angle=0,scale=0.36]{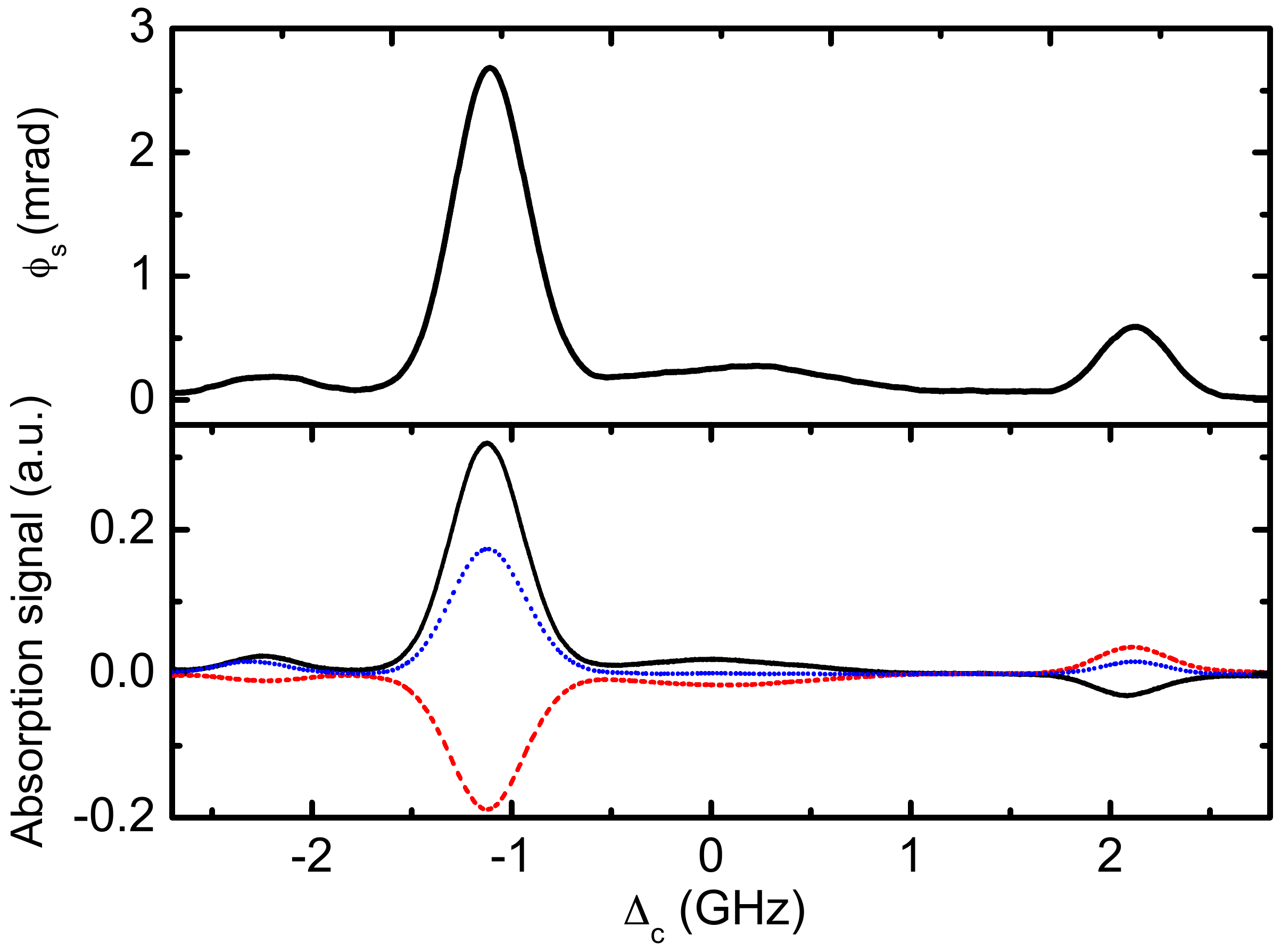}
\caption{(a) Measured phase shift of the probe using heterodyne detection technique. (b) Observed signal from direct probe absorption measurement with coupling being $\sigma^{+}$ and probe being $\sigma^{-}$. The solid and the dashed curves correspond to the absorption of linear polarization component in vertical and horizontal directions respectively. The dotted curve correspond to the total absorption of the probe light.}
\label{fig5}
\end{figure}

In a further study of comparing this technique with the direct absorption measurement of the probe, the imaginary part of susceptibility can be calculated as $\Im\left(\chi_{3L}\right)=\frac{\Omega_{eff}}{\sqrt{\pi}\Delta kv_p}\frac{N\left|\mu_{eg}\right|^2}{\epsilon_0\hbar}\left(\frac{\Gamma_{rg}}{\Omega_p^2}\right)\rho_{rr}$. The imaginary part is linked with the real part of susceptibility as $\Im(\chi_{3L})=\frac{2\Gamma_{rg}\Delta_{p}}{\Omega^{2}_{p}}\Re(\chi_{3L})$. If $\Delta_{p}=1$ GHz and $\Omega_{p}=250$ MHz, then the phase shift, ($k_{p}l\Re(\chi_{3L})$) becomes two orders of magnitude larger compared to the absorption ($k_{p}l\Im(\chi_{3L})$). In the experiment, direct absorption of the probe beam was measured with a improved sensitivity of the lock-in amplifier. The comparison of the direct probe absorption measurement with heterodyne detection technique is presented in figure~\ref{fig5}. It has been observed that the direct probe absorption is very sensitive to the small change in the polarizations of the coupling and the probe beams. As shown in figure~\ref{fig5}, the absorption of the orthogonal linear polarization components of the $\sigma^{-}$ light are opposite in sign and the total absorption is shown to be less compared to one of the linear polarization component.  Also, the signs of the absorption signals corresponding to different hyperfine states of the same spectrum are opposite to each other. The behaviour of the absorption spectrum observed from direct probe absorption measurement can only be explained by the small polarization rotation of the probe field due to the impurity in the circular polarization of the coupling laser beam. Considering, $\rho_{rr}$ being $10^{-5}$, the probe absorption will be of the order of $10^{-7}$ for the given laser parameters and such a small absorption can easily be obscured in the polarization rotation of the probe due their small polarization impurity due to non-ideal optical components. Hence, direct probe absorption measurement in this case can't provide a reliable measurement of the Rydberg population for very small absorption. In contrast, the signal observed in heterodyne detection technique is only sensitive to the relative phase between probe and the reference beams, but is insensitive to any small change in their power or polarization. Hence, the observed dispersion signal giving the information about Rydberg population becomes robust and reliable.   

\section{Conclusion}
In conclusion, we have demonstrated an all optical technique to measure the two-photon excitation to Rydberg state. The technique is found to be sensitive enough to measure Rydberg population of the order of $10^{-7}$. With  such a small Rydberg population, direct absorption measurement can't be reliable since it will be obscured by small polarization rotation due to the imperfection in the polarization components of light used in the experiment. The technique is particularly useful in the experiments with Rydberg excitation in thermal vapour. However, the technique can also be used for ultra-cold atoms with suitable laser parameter.

\section*{Acknowledgement}
We acknowledge Sushree S. Sahoo and Snigdha S. Pati for assisting in performing the experiment. This experiment was financially supported by the Department of Atomic Energy, Govt. of India.




\begin{thebibliography}{99}
\bibitem{rabi}The probe Rabi frequency is estimated using the relation with the intensity of probe as $I/I_{sat}=2\left|\Omega_p\right|^2/\Gamma^2$. For $^{85}$Rb, $I_{sat}=1.64$ mW and $\Gamma=2\pi\times6$ MHz. Coupling Rabi frequency is estimated by fitting the Rydberg EIT signal with a weak probe beam without focusing and then scaling it with the intensity of the beam.
\end{thebibliography}
\end{document}